%% file: arXiv14_sec-inv.tex
\documentclass[10pt,conference]{IEEEtran}
\usepackage{amsmath, amssymb, url}
\usepackage{amsfonts, amsbsy}
\usepackage{fixltx2e}
\usepackage{graphicx}
\usepackage[usenames]{color}

\newtheorem{assumption}{Assumption}
\newtheorem{theorem}{Theorem}
\newtheorem{proposition}{Proposition}

\ifodd 1
\newcommand{\com}[1]{\textbf{\color{red}(COMMENT: #1)}} %comment of the text
\newcommand{\comm}[1]{\textbf{\color{green}(#1)}} %comment of the text
\newcommand{\clar}[1]{\textbf{\color{green}(NEED CLARIFICATION: #1)}}

\else

\newcommand{\com}[1]{}
\newcommand{\comm}[1]{}
\newcommand{\clar}[1]{}
\fi
\begin{document}

\title{Closing the Price of Anarchy Gap in the Interdependent Security Game}

\author{\authorblockN{Parinaz Naghizadeh and Mingyan Liu}
\authorblockA{Department of Electrical Engineering and Computer Science\\
University of Michigan, Ann Arbor, Michigan, 48109-2122\\
Email: \{naghizad, mingyan\}@umich.edu}}
%\and
%\authorblockN{Mingyan Liu}
%\authorblockA{Department of Electrical Engineering and Computer Science\\
%University of Michigan, Ann Arbor, Michigan, 48109-2122\\
%Email: mingyan@umich.edu}}

\maketitle

\begin{abstract}

The reliability and security of a user in an interconnected system depends on all users' collective effort in security. Consequently, investments in security technologies by strategic users is typically modeled as a public good problem, known as the Interdependent Security (IDS) game. The equilibria for such games are often inefficient, as selfish users free-ride on positive externalities of others' contributions. In this paper, we present a mechanism that implements the socially optimal equilibrium in an IDS game through a message exchange process, in which users submit proposals about the security investment and tax/price profiles of one another. This mechanism is different from existing solutions in that (1) it results in socially optimal levels of investment, closing the Price of Anarchy gap in the IDS game, (2) it is applicable to a general model of user interdependencies.  We further consider the issue of individual rationality, often a trivial condition to satisfy in many resource allocation problems, and argue that with positive externality, the incentive to stay out and free-ride on others' investment can make individual rationality much harder to satisfy in designing a mechanism. 

\end{abstract}

\input{Intro}
\input{Model}

\input{Mechanism}
\input{Individual}

\input{Conclusion}

\input{Appendix}

\bibliographystyle{IEEEtran}
\bibliography{IEEEabrv,sec_inv}

\end{document}

%% file: Intro.tex
\section{Introduction} \label{sec:intro}

As a result of the rapid growth of the Internet, networks of all kinds, and file sharing systems, the security of a user, or entity, or a network\footnote{These terms are used interchangeably in this paper to denote a single unit in a connected system.}, in the context of a bigger system of connected users, entities or networks, is no longer solely determined by that user's own investment in security, but becomes increasingly dependent on the effort exerted by the collection of interconnected users. Accordingly, the security and reliability of the interconnected system is viewed as a {\em public good}, for which the investments in security exhibit a \emph{positive externality} effect: the investment of one user on security technologies will also improve the security posture of the other users interacting with it. Consequently, strategic users can choose to free-ride on others' effort, resulting in an overall under-investment in security. 

This problem of (under-)investment in security by an interconnected group of selfish users, both in general as well as in the context of computer security, has been extensively studied in the framework of game theory, see e.g. \cite{kun03, varian04, parameswaran07, grossklags08, grossklags10, lelarge09, walrand11, laszka12}, and is often referred to as the Interdependent Security (IDS) game. IDS games were first presented by Kunreuther and Heal \cite{kun03} to study the incentive of airlines to invest in baggage checking systems, and by Varian \cite{varian04} in the context of computer system reliability. In the majority of these papers, under-investment in security is verified by finding the levels of effort exerted in a Nash equilibrium of the IDS game, and comparing them with the socially optimal levels of investment. 

The increasing number of unprotected devices connected to the Internet, the constant emergence of new security threats, and the insufficiency of improved security technologies in compensating for the under-investment problem \cite{walrand11}, motivates the study of mechanisms for improving network security. Several methods for increasing users' investments, and thus the reliability of the interconnected system, have been proposed in the literature. These mechanisms fall into two main categories, based on whether they \emph{incentivize} or \emph{dictate} user cooperation. Mechanisms that dictate user investment in security, e.g. regulations, audits, and third party inspections \cite{laszka12}, leverage the power of an authority such as the government or an Internet service provider (ISP). These methods are only effective if the authority has enough power to accurately monitor users and establish a credible threat of punishment. 

Among the mechanisms that incentivize user investment in security, \emph{cyberinsurance} is one of the most commonly studied approaches \cite{kun03, pal10, laszka12}. Using insurance, users transfer part of the security risks to an insurer in return for paying a premium fee. Cyberinsurance is affected by the classic insurance problems of adverse selection (higher risk users seek more protection) and moral hazard (users lower their investment in self-protection after being insured). Therefore, the insurance company needs to somehow mitigate the information asymmetry and calculate the premium fees with these considerations in mind. {An example of such solutions is when an insurer chooses to monitor investments and/or inspect users' devices to prevent the moral hazard problem, specifying the terms of the contract accordingly to ensure appropriate levels of investment in self-protection \cite{kun03}.  
}

A method similar to insurance is proposed in \cite{parameswaran07}, where a certifying authority classifies users based on whether or not they have made security investments, and ensures that certified users get adequate compensation in case of a security incident. Another theoretically attractive incentive mechanism that may result in optimal levels of investment is the {\em liability rule} \cite{kun03, varian04}, where users are required to compensate others for the damages caused by their under-investment in security. However, these mechanisms are costly in that it is difficult to accurately determine the cause of a damage. Alternatively, \cite{varian04} proposes assigning a level of \emph{due care}, in which following a security incident, a user is penalized only if its level of investment is lower than a pre-specified threshold. Finally, users can be incentivized to invest in security if they are assigned bonuses/penalties based on their security outcome (e.g. users get a reward if their security has not been breached), or get subsidized/fined based on their effort (e.g. users are given discounts if they buy security products) \cite{grossklags10}. 

%It should be noted that in all the aforementioned incentive mechanisms, there is a need for either auditing users, monitoring their actual investment, or accurately observing their security outcome. In addition to incurring extra costs, such monitoring may cause privacy concerns. 

In this paper, we take a mechanism design approach to the security investment problem. Specifically, we present a game form, 
consisting of a message exchange process and an outcome function, through which users converge to an equilibrium where they make the socially optimal levels of investment in security. Our method is different from the previous solutions in several ways, highlighted as follows. 
\begin{enumerate}
\item The proposed mechanism is applicable to the general model of interdependence proposed in \cite{walrand11}. This model allows continuous levels of effort (as opposed to a binary decision of whether or not to invest in security \cite{kun03, parameswaran07, lelarge09}).  
\item It does not assume perfect protection once investment is made (unlike epidemic models \cite{kun03, laszka12}). {Another similar assumption is to decompose the risks of a user into direct and indirect (i.e. spreading from another infected user) risks, and assume perfect protection against direct risks only \cite{laszka12}. Nevertheless, none of these models can be descriptive of an IDS game, as no security technology can provide perfect protection against all threads. }  

\item It models the heterogeneity in users' preferences and their importance to the system by allowing for a more general utility function (in contrast to \cite{kun03, varian04, grossklags08, grossklags10, pal10, lelarge09}). 
\item This mechanism not only improves the levels of investment (as also done in \cite{walrand11}), but in fact results in socially optimal investments in security. 
%\item More importantly, in the proposed mechanism, once initial participation is ensured, users are fully incentivized to invest in the optimal levels of security without any need for monitoring. Therefore, the mechanism preserves the privacy of autonomous entities, since there is no need to audit users, monitor their actions, or dictate the outcome.  
\end{enumerate}

The rest of this paper is organized as follows. In Section \ref{sec:model}, we present a model for the IDS game. {We introduce the concept of price of anarchy in Section \ref{sec:PoA}, and highlight the inefficiency of Nash equilibria in an unregulated IDS game through a simple example.} We discuss the decentralized mechanism and its optimality in Section \ref{sec:mec}. Section \ref{sec:IR} illustrates that such optimal mechanism may fail to be individually rational, typically a trivial requirement in many other settings. Section \ref{sec:conclusion} concludes the paper with directions for future work. 

%% file: Model.tex
\section{Model and Preliminaries} \label{sec:model}

Consider a collection of $N$ users; this collection will also be referred to as the system. Each user $i$ can choose a level $x_i\geq 0$ of effort/investment in security, incurring a cost $c_i>0$ per unit of investment. Let $\mathbf{x} = \{x_1, x_2, \ldots, x_N\}$ denote the vector of investments. A user $i$'s \emph{security risk} function is denoted by $f_i(\mathbf{x})$. The security risk function models the expected losses of an individual in case of a security breach. These functions vary among users depending on both their security interdependencies and their valuations of security. We make the following assumptions about the functions $f_i(\cdot)$: 

\begin{assumption}
$f_i(\cdot) > 0$ is differentiable and decreasing in $x_j$, for all $i$ and all $j$.
\end{assumption}
The assumption of $\partial f_i/\partial x_j < 0$ models the positive externalities of security investments. 
 
\begin{assumption}
$f_i(\cdot)$ is strictly convex.
\end{assumption}
The assumption of convexity means that initial investment in security offers considerable protection to the users \cite{laszka12, top35}. However, even with high effort, it is difficult to reduce the cost to zero, as there is no strategy that could prevent all malicious activity \cite{walrand11, top35}. 

\vspace{\baselineskip}

The utility function of a user $i$ is defined as:
\begin{eqnarray}
u_i(\mathbf{x}) = -f_i(\mathbf{x}) - c_ix_i - t_i ~.
\label{eq:utility}
\end{eqnarray}
In \eqref{eq:utility}, $g_i(\mathbf{x}) := f_i(\mathbf{x}) + c_ix_i$ is referred to as the \emph{cost function} of user $i$ \cite{walrand11}, and represents all the costs associated with security investments and breaches. The term $t_i$ is the \emph{monetary transfer} that can be imposed on/awarded to users throughout the mechanism, which may itself depend on the vector of investments $\mathbf{x}$ (as detailed shortly).  This term is commonly known as {\em numeraire commodity} in the literature of mechanism design \cite{mas95}, as opposed to the {\em commodity of interest}, which are the security investments in our context. {To illustrate the purpose of including this term in a user's utility function, note that externalities are defined as the side-effects of users' actions on one another, the costs or benefits of which are not accounted for when users pick their actions. A numeraire commodity is often used in problems involving externalities to bring such side-effects into strategic individuals' decision making process, a tactic referred to as ``internalizing the externalities''. } 

We make the following assumptions about the users:
\begin{assumption}
All users $i$ are strategic, and choose their investment $x_i$ in order to maximize their own utility function \eqref{eq:utility}. 
\end{assumption}

\begin{assumption}
The cost $c_i$ and the functional form of $f_i(\cdot)$ are user $i$'s private information. 
\end{assumption}

\vspace{\baselineskip}

The Interdependent Security (IDS) game induced among these $N$ strategic players is defined as the strategic game $(\{1,\ldots, N\}, \{x_i\geq 0\}, \{u_i(\cdot)\})$. 
The \emph{socially optimal} vector of security investments {in this $N$ user system} is the vector $\mathbf{x}^*$ maximizing the social welfare, as determined by the solution to the following centralized problem:  
\begin{eqnarray}
& \max\limits_{\substack{(\mathbf{x},\mathbf{t})}}& \quad \sum_{i=1}^N u_i(\mathbf{x}) \notag\\
&\text{s.t.}& \quad \sum_{i=1}^N t_i = 0~, \quad \mathbf{x}\succeq 0 ~. \notag\\
\equiv &\min\limits_{\substack{\mathbf{x}}}& \quad \sum_{i=1}^N g_i(\mathbf{x}) \notag\\
&\text{s.t.}& \quad \mathbf{x}\succeq 0 ~.
\label{eq:Pcd}
\end{eqnarray}
In other words, socially optimal solutions minimize the social cost $G(\mathbf{x}) := \sum_{i=1}^N g_i(\mathbf{x})$. By Assumption 2, there is a unique socially optimal investment profile $\mathbf{x}^*$ for Problem \eqref{eq:Pcd}.  Also, due to Assumptions 3 and 4, there is no individual/user in the system with enough information to determine $\mathbf{x}^*$. 

Accordingly, our goal is to find a mechanism, run by a  manager/regulator, {such that the induced interdependent security game has as its equilibrium the solution to the centralized problem \eqref{eq:Pcd} (also referred to as ``implementing'' the solution to \eqref{eq:Pcd}).}  
 
To determine the effort that users exert in an IDS game, with or without regulation (i.e., $t_i=0, \ \forall i$), we will consider the vector of investments $\mathbf{x}$ in a Nash equilibrium (NE) of the game $(\{1,\ldots, N\}, \{x_i\geq 0\}, \{u_i(\cdot)\})$. Theoretically, Nash equilibria describe users' actions in a game of complete information. However, due to Assumption 4, the model studied herein is one of incomplete information. The Nash equilibrium in this game can be interpreted as the convergence point of an iterative process, in which each user adjusts its action at each round based on its observations of other users' actions, until unilateral deviations are no longer profitable \cite{sharma11, walrand11}.\footnote{Alternatively, one may relax Assumption 4 and study a game of complete information, as is done in the majority of the current literature on IDS games.}

A pure strategy Nash equilibrium of the IDS game is a vector of investments ${\mathbf{\bar{x}}}$, for which,   
\begin{eqnarray}
u_i(\bar{x}_i, \mathbf{\bar{x}}_{-i}) \geq u_i({x}_i, {\mathbf{\bar{x}}}_{-i}), \qquad \forall x_i\geq 0,\ \forall i~.
\label{eq:NE-def}
\end{eqnarray} 

We first ensure that the game studied indeed has a Nash equilibrium in the following result.  The proof can be found in the Appendix. 
\begin{proposition}\label{prop1}
There always exists a pure strategy Nash equilibrium in an unregulated (i.e. $t_i=0, \ \forall i$) IDS game modeled in this section.  
\end{proposition}

%%%%%%%%%%%%%%%%%%%%%%%%%%%%%%%%%%%%%%%%%%%%%%%%%%%%%%%%%%%%%%%%%%%%%%%%%%%%%%%%%%%%%%%%%%% 
\section{Price of Anarchy in an Unregulated IDS Game} \label{sec:PoA}

Existence notwithstanding, the Nash equilibria of an unregulated IDS game are often inefficient.  A common metric for quantifying the inefficiency of such equilibria is the \emph{Price of Anarchy} (PoA), defined as the largest possible ratio between the worst possible social cost at a Nash equilibrium $\mathbf{\bar{x}}$ and at the social optimum $\mathbf{x}^*$. Formally, PoA $\rho$ is defined as: 
\begin{eqnarray}
\rho &=& \max_{\mathbf{\bar{x}}} \rho(\mathbf{\bar{x}})~, \nonumber \\
\rho(\mathbf{\bar{x}})&:=& \frac{G(\mathbf{\bar{x}})}{G(\mathbf{x}^*)} = \frac{\sum_{i=1}^N g_i(\mathbf{\bar{x}})}{\sum_{i=1}^N g_i(\mathbf{x}^*)}~.
\label{eq:PoA-def}
\end{eqnarray} 

In \cite{walrand11}, the authors characterize the price of anarchy in an {unregulated} IDS game, i.e., the game in which no external mechanism is implemented. The NE of this game is defined in the same way as in \eqref{eq:NE-def}, with $u_i(\cdot)$ replaced by $-g_i(\cdot)$. This means that without regulation, users selfishly pick effort levels that minimize their own cost. As a result, $\rho>1$ for several plausible risk function models (\cite[Lemma 1, Propositions 2, 3]{walrand11}, reflecting under-investment in security.  {Below we present such an example, different from the aforementioned results presented in \cite{walrand11}, and chosen for its simplicity.}

Consider $N$ interconnected users, and a \emph{total effort} model for users' risk function, such that  
\begin{eqnarray*}
f_i(\mathbf{x}) = f(\sum_{j=1}^N x_j),\ \forall i.
\end{eqnarray*} 
Furthermore, without loss of generality, assume $c_1<c_2<\cdots<c_N$. At the Nash equilibrium of this game, each user will choose a level of investment $x_i\geq 0$ to minimize its own cost. Therefore, at the Nash equilibrium $\mathbf{\bar{x}}$ we must have: 
%\[\begin{array}{ll}
\begin{eqnarray*}
		\bar{x}_i = 0 & \text{if }& \frac{\partial f(\bar{x}_i, \mathbf{\bar{x}}_{-i})}{\partial x_i} + c_i > 0~, \\
		\bar{x}_i > 0 & \text{if }& \frac{\partial f(\bar{x}_i, \mathbf{\bar{x}}_{-i})}{\partial x_i} + c_i = 0~.
\end{eqnarray*}
%\end{array}\]
We conclude that only the user with the lowest cost will be exerting a non-zero effort at the Nash equilibrium $\mathbf{\bar{x}}$. Thus: 
\[\partial f(\bar{x}_1, \mathbf{0})/ \partial x_1 = - c_1, \text{  and } \bar{x}_j = 0, \ \forall j>1~.\] 

At the socially optimal equilibrium $\mathbf{x}^*$ on the other hand, the levels of investment are determined by: 
\begin{eqnarray*}
		{x}_i^* = 0 & \text{if } & N\ \frac{\partial f({x}^*_i, \mathbf{{x}}^*_{-i})}{\partial x_i} + c_i > 0~, \\
		{x}_i^* > 0 & \text{if } & N\ \frac{\partial f({x}^*_i, \mathbf{{x}}^*_{-i})}{\partial x_i} + c_i = 0~.
\end{eqnarray*}
Again the user with the lowest cost will be exerting all the effort at the equilibrium $\mathbf{{x}}^*$, however at a higher level, determined by:  
\[\partial f({x}^*_1, \mathbf{0})/ \partial x_1 = -c_1/N, \text{  and } {x}^*_j = 0, \ \forall j>1~.\]

\begin{figure}%
\centering
\includegraphics[width=\columnwidth]{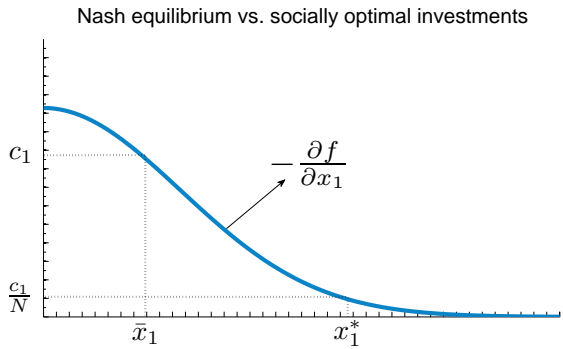}%
\caption{Under-investment in security in an unregulated IDS game.}%
\label{NEvsSO}%
\end{figure}

The price of anarchy will therefore be given by: 
\[\rho = \frac{N\ f(\bar{x}_1, \mathbf{0}) + c_1\bar{x}_1}{N\ f({x}^*_1, \mathbf{0}) + c_1{x^*_1}}~.\]
By the strict convexity of $f(\cdot)$, we have:
\[f(\bar{x}_1, \mathbf{0}) - f({x}^*_1, \mathbf{0}) > \frac{\partial f({x}^*_1, \mathbf{0})}{\partial x_1} (\bar{x}_1 - x^*_1)~.\]
Hence, $\rho>1$. 
{Figure \ref{NEvsSO} illustrates the levels of investment in both the socially optimal and the Nash equilibrium of this game. Based on fig. \ref{NEvsSO}, it is easy to observe the under-investment in security in the Nash equilibrium of an unregulated game.}  

In the next section, we present a mechanism under which all Nash equilibria of the induced IDS game coincide with the socially optimal solution, i.e., we will have $\rho = 1$, closing the price of anarchy gap.

%% file: Mechanism.tex
\section{A Positive Externality Security Investment Mechanism (PESIM)} \label{sec:mec} 

In this section, we present a mechanism that implements the socially optimal solution to \eqref{eq:Pcd} in an informationally decentralized setting. This mechanism is adapted from \cite{sharma11, hurwicz79}. 

%%%%%%%%%%%%%%%%%%%%%%%%%%%%%%%%%%%%%%%%%%%%%%%%%%%%%%%%%%%%%%%%%%%%%%%%%%%%%%%%%%
%\subsection{The Game Form} 
A decentralized mechanism is specified by a game form $({\cal M}, h)$. 
\begin{itemize}
\item The message space ${\cal M}:=\Pi_{i=1}^N {\cal M}_i$ specifies the set of permissible messages ${\cal M}_i$ for each user $i$. 
\item The outcome function $h:{\cal M}\rightarrow {\cal A}$ determines the outcome of the game based on the users' messages. Here, ${\cal A}$ is the space of all security investment profiles and tax profiles, i.e., $(\mathbf{x}, \mathbf{t})$.  
\end{itemize}
The game form, together with the utility functions, define a game, represented by $({\cal M}, h(\cdot), \{u_i(\cdot)\})$. {This will also be referred to as the regulated IDS game.} 

We say the message profile $\mathbf{{m}^*}$ is a Nash equilibrium of this game, if  
\begin{eqnarray}
u_i(h({m}^*_i, \mathbf{{m}^*}_{-i})) \geq u_i(h({m}_i, {\mathbf{{m}^*}}_{-i})), \qquad \forall m_i,\ \forall i~.
\label{eq:NE-def-2}
\end{eqnarray} 

The components of the proposed decentralized PESIM mechanism are specified as follows. 

\textbf{The Message Space:} 
Each user $i$ reports a message $m_i:=(\boldsymbol{\pi}_i, \mathbf{x}_i)$ to the regulator, with $\boldsymbol{\pi}_i\in \mathbb{R}^N_{+}$ and $\mathbf{x}_i\in \mathbb{R}^N$. The component $\mathbf{x}_i$ is user $i$'s proposal regarding the public good, i.e., the security investment profile, while $\boldsymbol{\pi}_i$ is user $i$'s suggestion regarding the private good, i.e., the price profile\footnote{{Note the use of the term \emph{price profile} for the vectors $\boldsymbol{\pi}_i$. As illustrated later, these terms are closely related to Lindhal prices, and will in turn be used to determine a \emph{tax profile} $\mathbf{t}$. } }.

\textbf{The Outcome Function:} 
The outcome function $h$ takes the message profile $\mathbf{m}$ as input and determines the security investment profile $\mathbf{\hat{x}}$ and the tax profile $\mathbf{\hat{t}}$ as follows: 
\begin{eqnarray}
\mathbf{\hat{x}}(\mathbf{m}) &=& \frac{1}{N} \sum_{i=1}^N \mathbf{x}_i~, \label{eq:out_x}\\
\mathbf{\hat{t}}_i(\mathbf{m}) &=& (\boldsymbol{\pi}_{i+1} - \boldsymbol{\pi}_{i+2})^T\mathbf{\hat{x}}(\mathbf{m})\notag\\
&+& (\mathbf{x}_{i} - \mathbf{x}_{i+1})^T \text{diag}(\boldsymbol{\pi}_i)(\mathbf{x}_{i} - \mathbf{x}_{i+1})\notag\\
&-& (\mathbf{x}_{i+1} - \mathbf{x}_{i+2})^T \text{diag}(\boldsymbol{\pi}_{i+1})(\mathbf{x}_{i+1} - \mathbf{x}_{i+2}), \forall i.~~~~  
\label{eq:out_t}
\end{eqnarray}
In \eqref{eq:out_t}, for simplicity $N+1$ and $N+2$ are treated as $1$ and $2$, respectively.  {That is, $N+1$ denotes the modulo $(N \mbox{ mod } 1)$, and so on.} 

This outcome function is interpreted as follows: first, \eqref{eq:out_x} states that the contribution $\hat{x}_i$ of each user $i$ to the public good vector of investments $\mathbf{\hat{x}}$ is determined by the average of all users' proposals. The taxation term \eqref{eq:out_t} is then used to make sure that all investment profile proposals $\mathbf{x}_i$ are the same at equilibrium, and are equal to the socially optimal security investments. 

The tax term for user $i$ itself consists of three different terms. The first term $(\boldsymbol{\pi}_{i+1} - \boldsymbol{\pi}_{i+2})^T\mathbf{\hat{x}}(\mathbf{m})$ is independent of user $i$'s proposal for prices, and depends only on the investment profile\footnote{$\boldsymbol{\pi}_{i+1} - \boldsymbol{\pi}_{i+2}$ is interpreted as the Lindhal price for the public good \cite{hurwicz79}.}. The second term determines the penalties for the discrepancy between user $i$'s proposal $\mathbf{x}_i$ and user $(i+1)$'s proposal. This term will ensure eventual agreement between investment proposals put forward by different users.  The third term does not depend on user $i$'s message, and is used only as a balancing term. In fact, at equilibrium, both the second and third terms will be equal to zero. Nevertheless, their inclusion is necessary to ensure convergence to the optimal security investment profile, and also for budget balance (i.e., the sum of all taxes equal zero) on and off the equilibrium. Note that having budget balance off equilibrium is an important property of the proposed mechanism, in order to prevent complications in an iterative message exchange process that leads to the desired Nash equilibrium.

{We would also like to highlight the close relation between the tax term proposed in \eqref{eq:out_t} and the positive externalities of users' actions. As illustrated later, at an equilibrium of the PESIM mechanism, the second and third terms in \eqref{eq:out_t} disappear, so that the tax $\mathbf{\hat{t}}_i$ for user $i$ reduces to $\mathbf{\hat{t}_i} = {\mathbf{l}_i^*}^T\mathbf{\hat{x}}$, where $\mathbf{l}_i^*:= \boldsymbol{\pi}^*_{i+1} - \boldsymbol{\pi}^*_{i+2}$ is known as the Lindhal price for user $i$. Furthermore, when users' monetary taxes are assessed according to Lindhal prices, the socially optimal investments $\mathbf{x}^*$ will be individually optimal as well, i.e.,\footnote{See proof of Theorem \ref{th1} presented later in this section for the derivation of this result.} 
\begin{eqnarray}
\mathbf{x}^* = \arg\min_{\mathbf{x}\succeq 0}\quad g_i(\mathbf{x})+{\mathbf{l}_i^*}^T \mathbf{x}~. 
\end{eqnarray}  
As a result, it is easy to show that for all $i$, and all $j$ for which $\hat{x}_j\neq 0$, 
\begin{eqnarray}
\frac{\partial g_i(\hat{\mathbf{x}})}{\partial x_j}<0 \Rightarrow {\mathbf{l}_i^*}_j >0 \Rightarrow {\mathbf{l}_i^*}_j \hat{x}_j^* > 0~. \label{eq:int1}
%\mbox{if}\quad \frac{\partial g_i(\hat{\mathbf{x}})}{\partial x_j}>0 \Rightarrow {\mathbf{l}_i^*}_j <0 \Rightarrow {\mathbf{l}_i^*}_j \hat{x}_j^* < 0~.
\end{eqnarray} 
The interpretation of this observation is that by implementing the PESIM mechanism, user $i$ will be paying a monetary tax to user $j$, which is proportional to the positive externality of $j$'s investment on user $i$'s costs \eqref{eq:int1}. }

It should be pointed out that for the time being, we have assumed users' participation in the mechanism is ensured, either through policy mandate (e.g., the government may require users to participate in the mechanism as a prerequisite for conducting business with it), or secondary financial incentive (e.g., product discount for joining the collection of users interested in the mechanism), such that the incentive for participation is separate from the mechanism itself. In Section \ref{sec:IR}, we present a counter-example to illustrate why the individual rationality constraint, i.e., the condition that a user is better off by participating than staying out, may fail to hold, and discuss some implications of this observation. 

We close this section by presenting the theorems that establish the optimality of the proposed game form. Note that to prove this optimality, we first need to show that a profile $(\mathbf{\hat{x}}(\mathbf{{m}^*}),\mathbf{\hat{t}}(\mathbf{{m}^*}))$, derived at the NE $\mathbf{{m}^*}$ of the induced game, is an optimal solution to the centralized problem \eqref{eq:Pcd}, and therefore socially optimal. Furthermore, as the procedure for convergence to NE is not specified, we need to verify that the {optimality} property holds \emph{for all Nash equilibrium} of the message exchange process. This guarantees that the outcome will converge to the socially optimal solution regardless of the realized NE. These two requirements are established in Theorem \ref{th1} below. 

\begin{theorem}
Let  $(\mathbf{\hat{x}}(\mathbf{{m}^*}),\mathbf{\hat{t}}(\mathbf{{m}^*}))$ be the investment and tax profiles obtained at the Nash equilibrium $\mathbf{{m}^*}$ of the game $({\cal M}, h(\cdot), \{u_i(\cdot)\})$. Then, $(\mathbf{\hat{x}},\mathbf{\hat{t}})$ is an optimal solution to the centralized problem \eqref{eq:Pcd}. Furthermore, if $\mathbf{\bar{m}}$ is any other Nash equilibrium of the proposed game, then $\mathbf{\hat{x}}(\mathbf{\bar{m}}) = \mathbf{\hat{x}}(\mathbf{{m}^*})$. 
\label{th1}
\end{theorem} 

{\em Proof:} Let $\mathbf{{m}^*}$ be a Nash equilibrium of the message exchange process, resulting in an allocation $(\mathbf{\hat{x}}, \mathbf{\hat{t}})$. Assume user $i$ updates its message from ${m}^*_i = (\boldsymbol{\pi}^*_i, \mathbf{x}^*_i)$ to ${m}_i = (\boldsymbol{{\pi}}_i, \mathbf{x}^*_i)$, that is, it only updates the price vector proposal. Therefore, according to \eqref{eq:out_x}, $\mathbf{\hat{x}}$ will remain fixed, while based on \eqref{eq:out_t}, the second term in $\hat{t}_i$ will change. Since  $\mathbf{{m}^*}$ is an NE, unilateral deviations are not profitable. Mathematically, 
\begin{eqnarray}
&& (\mathbf{x}^*_{i} - \mathbf{x}^*_{i+1})^T \text{diag}(\boldsymbol{\pi}^*_i)(\mathbf{x}^*_{i} - \mathbf{x}^*_{i+1})\notag\\
&\leq& (\mathbf{x}^*_{i} - \mathbf{x}^*_{i+1})^T \text{diag}(\boldsymbol{\pi}_i)(\mathbf{{x}}^*_{i} - \mathbf{{x}}^*_{i+1}), \quad \forall \boldsymbol{\pi}_i \succeq 0~.~~~
\label{eq:pf1-1}
\end{eqnarray} 
Hence, from \eqref{eq:pf1-1} we conclude that for all $i$: 
\begin{eqnarray}
\mathbf{x}^*_{i} = \mathbf{x}^*_{i+1} \quad \text{or} \quad \boldsymbol{\pi}^*_i = \mathbf{0}~.
\label{eq:pf1-2}
\end{eqnarray}
Using \eqref{eq:pf1-2} together with \eqref{eq:out_t} we conclude that at equilibrium, the second and third terms of a user's tax vanish. Denoting $\mathbf{l}_i^*:= \boldsymbol{\pi}^*_{i+1} - \boldsymbol{\pi}^*_{i+2}$, we get: 
\begin{eqnarray}
\mathbf{\hat{t}_i}(\mathbf{m}^*) = {\mathbf{l}_i^*}^T\mathbf{\hat{x}}(\mathbf{m}^*)~.
\label{eq:pf1-3}
\end{eqnarray}

Now consider the utility function of the users at the Nash equilibrium $\mathbf{m}^*$. Since unilateral deviations are not profitable, a user's utility \eqref{eq:utility} should be maximized at the NE, i.e., for any choice of $\mathbf{x}_i$ and $\boldsymbol{\pi}_i \succeq 0$: 
\begin{eqnarray}
&& g_i(\mathbf{\hat{x}}(\mathbf{m}^*)) + {\mathbf{l}_i^*}^T\mathbf{\hat{x}}(\mathbf{m}^*) \notag\\
&\leq& g_i(\frac{\mathbf{x}_i + \sum_{j\neq i} \mathbf{x}_j^*}{N}) + {\mathbf{l}_i^*}^T \frac{\mathbf{x}_i + \sum_{j\neq i} \mathbf{x}_j^*}{N} \notag\\
&& + (\mathbf{x}_{i} - \mathbf{x}^*_{i+1})^T \text{diag}(\boldsymbol{\pi}_i)(\mathbf{x}_{i} - \mathbf{x}^*_{i+1})
\label{eq:pf1-4}
\end{eqnarray}
If we choose $\boldsymbol{\pi}_i=\mathbf{0}$ and let $\mathbf{x}_i = N\cdot \mathbf{x} - \sum_{j\neq i} \mathbf{x}_j^*$, where $\mathbf{x}$ is any vector of security investments, we get: 
\begin{eqnarray}
g_i(\mathbf{\hat{x}}(\mathbf{m}^*)) + {\mathbf{l}_i^*}^T \mathbf{\hat{x}}(\mathbf{m}^*) 
\leq g_i(\mathbf{x}) + {\mathbf{l}_i^*}^T \mathbf{x}, \quad \forall \mathbf{x}~.
\label{eq:pf1-5}
\end{eqnarray}

To show that the Nash equilibrium $\mathbf{m}^*$ results in a socially optimal allocation, we sum up \eqref{eq:pf1-5} over all $i$, and use the fact that $\sum_i \mathbf{l}_i^* = \mathbf{0}$ to get: 
\begin{eqnarray}
\sum_{i=1}^N g_i(\mathbf{\hat{x}}(\mathbf{m}^*)) \leq \sum_{i=1}^N g_i(\mathbf{x}), \quad \forall \mathbf{x}~.
\label{eq:pf1-7}
\end{eqnarray} 
Therefore, $\mathbf{\hat{x}}(\mathbf{m}^*)$ is the optimal investment profile minimizing the social cost in problem \eqref{eq:Pcd}. Furthermore, any tax profile $\mathbf{t}$ satisfying the budget balance condition can be chosen as the tax profile in the optimal solution. Since the tax terms \eqref{eq:pf1-3} are balanced, we conclude that $(\mathbf{\hat{x}}(\mathbf{m}^*), \mathbf{\hat{t}}(\mathbf{m}^*))$ solves \eqref{eq:Pcd} and is therefore socially optimal. Finally, since our choice of the NE $\mathbf{m}^*$ has been arbitrary, the same proof holds for any other NE, and thus all NE of the mechanism result in the optimal solution to problem \eqref{eq:Pcd}. 
{\raggedleft \hfill ${\blacksquare}$} 

Finally, we establish the converse of this statement in Theorem \ref{th2}, i.e., given an optimal investment profile, there exists an NE of the proposed game which implements this solution. 

\begin{theorem}
Let $\mathbf{x}^*$ be the optimal investment profile in the solution to the centralized problem \eqref{eq:Pcd}. Then, there exists at least one Nash equilibrium $\mathbf{{m}^*}$ of the game $({\cal M}, h(\cdot), \{u_i(\cdot)\})$ such that $\mathbf{\hat{x}}(\mathbf{{m}^*})=\mathbf{x}^*$. 
\label{th2}
\end{theorem}

The proof of this theorem is given in the appendix. 

%%%%%%%%%%%%%%%%%%%%%%%%%%%%%%%%%%%%%%%%%%%%%%%%%%%%%%%%%%%%%%%%%%%%%%%%%%%%%%%
%\subsection{Discussion}

%% file: Individual.tex
\section{On Individual Rationality} \label{sec:IR}

Thus far, we have assumed user participation in the message exchange process is ensured using external incentive mechanisms. Alternatively, one could try to guarantee voluntary participation of strategic users by establishing that the so-called {\em individual rationality} condition is satisfied, i.e., users gain when participating in the mechanism as opposed to staying out. 

Whether a mechanism is individually rational depends on the structure of the game form, as well as the actions available to users when opting out. A common assumption in the majority of public good and resource allocation problems, including the prior work on the decentralized mechanism presented in Section \ref{sec:mec} (\cite{hurwicz79, sharma08, sharma11}), is that users will get a zero share (of the public good or allotted resources) when staying out. Following this assumption, \cite{hurwicz79, sharma08, sharma11} establish the individual rationality of the presented mechanism. However, a similar line of reasoning is not applicable to the current problem. 

The different nature of individual rationality in an IDS game can be intuitively explained as follows. By implementing a socially optimal equilibrium, (some) users will be required to increase their level of investment in security. In turn, the mechanism should either guarantee that these users enjoy a higher level of protection due to higher equilibrium investments from other participants, and/or are adequately compensated for their contribution by a monetary reward {(negative taxation)}. 
On the other hand, by staying out, a user can still enjoy the positive externalities of other users' investments (although these may be lower when the mechanism has partial coverage), choose its  optimal action accordingly, and possibly avoid taxation. Thus to establish individual rationality in such an IDS game is not nearly as trivial as in previous studies. 

Indeed, the following counter-example shows that the benefits of staying out can overthrow that of participation, making a user better off when acting as a ``loner''. 

Specifically, a loner is a user who refuses to participate in the mechanism, and later best-responds to the socially optimal strategy of the remaining $N-1$ users who did participate.  Arguably, these $N-1$ users could also revise their strategy (investments) in response to this loner's best-response, leading to a sequential game.  In this example we will compare the loner's utility in the socially optimal solution when participating in the mechanism, versus the utility at the outcome of the sequential game described above. 

Consider a collection of $N$ users. Without loss of generality, assume $c_1<c_2<\ldots<c_N$. Assume user 1 is contemplating whether to participate or remain a free agent. We further assume all users have the same risk function $f_i(\mathbf{x}) = \exp{(-\sum_{i=1}^N x_i)}$ (an instance of the total effort model \cite{varian04}). 

It is easy to show that at the socially optimal solution $\mathbf{x}^*$ to the $N$-player game, the user with the smallest cost would exert all the effort (see e.g. Section \ref{sec:PoA}, or \cite{varian04}), such that: 
\[\exp(-x_1^*) = c_1/N, \quad x_j^* = 0, \ \forall j>1~.\] 
By \eqref{eq:pf1-3} in the proof of Theorem 1, the tax for user 1 is given by:
\[t_1^* = {\mathbf{l}_1^*}^T \mathbf{x}^* = l_{11}^* x_1^*~.\] 
Re-writing \eqref{eq:pf1-5} in the proof of Theorem 1 as

\begin{eqnarray}
\mathbf{x}^* = \arg\min_{\mathbf{x}\succeq 0} g_1(\mathbf{x})+{\mathbf{l}_1^*}^T \mathbf{x}~,\notag  
\end{eqnarray}
and applying the KKT conditions, we conclude that: 
\[l_{11}^* + \frac{\partial g_1}{\partial x_1} (\mathbf{x}^*)= l_{11}^*  - \exp(-x_1^*) + c_1 = 0\]
\[ \Rightarrow l_{11}^* = -(1- \frac{1}{N}) c_1 ~ \Rightarrow ~ t_1^* = -(1-\frac1N)c_1x_1^*~.\]
As expected, user 1 is getting a reward in this mechanism. 

Now assume user 1 opts out of the decentralized mechanism. The remaining $N-1$ users choose their strategies assuming user 1 exerts an effort of $x_1$. Then, by the nature of the total effort game, the user with the smallest cost among these $N-1$ players will exert all the effort (if any) such that: 
\[\exp(-x_1 - \hat{x}_2) = c_2/(N-1), \quad \hat{x}_j = 0, \ \forall j>2~.\]
On the other hand, if user 1 is best responding to a choice of ${x}_2$, it chooses an effort according to:  
\[\exp(-\hat{x}_1 - {x}_2) = c_1~.\]
Combining the last two equations, at an equilibrium $\hat{\mathbf{x}}$ of the sequential game we have: 
\[\hat{x}_1 = \arg\min_{x_1\geq 0} \exp(-x_1 - \max\{-\ln\frac{c_2}{N-1} - x_1, 0\}) + c_1x_1~.\]
From the above, we conclude that if $-\ln\frac{c_2}{N-1}$ is large enough, that is, if without user 1's participation, user 2 will exert a sufficiently high effort, user 1 will choose to free-ride. Otherwise, it may again exert all the effort, in which case $\exp(-\hat{x}_1) = c_1$. 

Let us focus on this latter case. It is interesting to note that the overall level of security in the sequential game is lower than the coordinated socially optimal equilibrium. 

We compare user 1's utility under the two scenarios.  
\begin{eqnarray*}
u_1^{IN}(\mathbf{x}^*) &=& -\exp{(-x_1^*)} - c_1x_1^* + (1-\frac1N)c_1x_1^*~.\notag\\
u_1^{OUT}(\mathbf{\hat{x}}) &=& -\exp{(-\hat{x}_1)} - c_1\hat{x}_1.
%\label{eq:}
\end{eqnarray*}
Therefore, 
\begin{eqnarray}
u_1^{IN} - u_1^{OUT} &=& -(\exp{(-x_1^*)}-\exp{(-\hat{x}_1)})\notag\\
&& - c_1(x_1^*-\hat{x}_1) + (1-\frac1N)c_1x_1^*\notag\\
&=& -(\frac{c_1}{N} - c_1) - c_1 (-\ln \frac{c_1}{N} + \ln c_1)\notag\\
&& + (1-\frac1N)c_1 (-\ln \frac{c_1}{N})\notag\\
&=& \frac{c_1}{N}\bigg( (N-1)(1-\ln{c_1}) - {\ln N} \bigg). 
\label{eq:ut_diff}
\end{eqnarray} 
Based on \eqref{eq:ut_diff}, with any cost $c_1\geq \exp(1)$, user 1's utility will decrease when participating, indicating that in this case the decentralized mechanism fails to satisfy individual rationality. 

In light of the above observation, we conclude that although the proposed mechanism is incentive compatible and implements the socially optimal levels of investment in a Nash equilibrium, it fails to satisfy individual rationality in general.  It remains an interesting question whether there are other mechanisms which would satisfy all requirements simultaneously, or alternatively whether this is a more fundamental challenge in designing mechanisms for resource allocation with positive externalities.  The answer should shed light on questions such as whether security policies should be mandated (or alternatively incentivized), rather than being left to users' free will.

%% file: Conclusion.tex
\section{Conclusion} \label{sec:conclusion}

In this paper, we have presented a decentralized mechanism, through which we can find and implement the socially optimal levels of investment in security in an interdependent security game. This mechanism is especially attractive as it is applicable to a wide range of user preferences, operates without the need for collecting information about these preferences, and does not need to centrally dictate the socially optimal outcome. %Moreover, the proposed mechanism does not require any monitoring of users actions or an assessment of their security outcomes. This property follows from the fact that the equilibrium concept used here is that of a Nash equilibrium, and therefore unilateral deviations from equilibrium actions are not profitable. As a result, once the mechanism converges to an equilibrium, the regulator does not need to ensure that users are exerting the socially optimal levels of effort, as it is already optimal for them to do so. 
 We further consider the issue of individual rationality, often a trivial condition to satisfy in many resource allocation problems.  We provide a counter example under the proposed mechanism, and argue that with positive externality, the incentive to stay out and free-ride on others' investment can make individual rationality much harder to satisfy in designing a mechanism.

The study of IDS games in the current framework can be further continued in several directions. First, the procedure and conditions under which the message exchange process converges to a Nash equilibrium remains an open problem, and is an interesting direction of future study. Alternatively, one could switch focus to Bayesian Nash equilibrium as the solution concept for games of incomplete information, to better capture the uncertainty of users about their environment, including other users' valuations of security and the resources available to them. It is also interesting to study how the information obtained from alternative resources, e.g. IP blacklists, can help users attain a better understanding of their security risks and consequently make more effective investment decisions.

%% file: Appendix.tex
\section*{Appendix} \label{sec:app}

In this appendix, we present the proofs to Proposition \ref{prop1} and Theorem \ref{th2}. %, and present an IDS game for which the price of anarchy is greater than 1. 
The proof for Theorem \ref{th2} is technically similar to that presented in \cite{sharma11, hurwicz79}, and the proof of Proposition \ref{prop1} follows from \cite[Proposition 1]{walrand11}. %These results are included here for reference. 

%%%%%%%%%%%%%%%%%%%%%%%%%%%%%%%%%%%%%%%%%%%%%%%%%%%%%%%%%%%%%%%%%%%%%%%%%%%%%%%%%%%%%%%%%%%%%%%%%%%%%%%%
\subsection*{Proof of Proposition 1} 
We first show that the strategy space $x_i\in [0, \infty)$ of a user $i$ can be effectively reduced to a convex and compact set. 

Let $\text{BR}_i(\mathbf{x}_{-i})$ represent user $i$'s best response to the strategies $\mathbf{x}_{-i}\succeq 0$ of all the other users. Define $\hat{x}_i = \frac{f_i(\mathbf{0}) + \epsilon}{c_i}$, for some $\epsilon>0$. By assumption 2, the functions $f_i(\cdot)$ are convex, and thus: 
\begin{eqnarray}
f_i(0, \mathbf{x}_{-i}) - f_i(\hat{x}_i, \mathbf{x}_{-i}) &\geq& -\hat{x}_i \ \frac{\partial f_i(\hat{x}_i, \mathbf{x}_{-i})}{\partial x_i}\notag\\
&=& -\frac{f_i(\mathbf{0}) + \epsilon}{c_i}\ \frac{\partial f_i(\hat{x}_i, \mathbf{x}_{-i})}{\partial x_i}~.~~~
\label{eq:prop1-1}
\end{eqnarray}  
By assumption 1, $f_i(\hat{x}_i, \mathbf{x}_{-i})\geq 0$, and $f_i(0, \mathbf{x}_{-i})\leq f_i(\mathbf{0})$. Therefore, \eqref{eq:prop1-1} reduces to: 
\begin{eqnarray}
f_i(\mathbf{0}) &\geq&  -\frac{f_i(\mathbf{0}) + \epsilon}{c_i}\ \frac{\partial f_i(\hat{x}_i, \mathbf{x}_{-i})}{\partial x_i}~.
\label{eq:prop1-2}
\end{eqnarray}  
Equation \eqref{eq:prop1-2} in turn implies that $\frac{\partial f_i(\hat{x}_i, \mathbf{x}_{-i})}{\partial x_i} + c_i > 0$. Therefore, since user $i$'s cost is increasing at $\hat{x}_i$, a best response to minimize the cost should be such that $\text{BR}_i(\mathbf{x}_{-i})\in [0, \hat{x}_i]$. Let $x_{max} := \max_{i} \hat{x}_i$. We conclude that for all $i$, the strategy sets can be effectively reduced to $x_i\in [0, x_{max}]$. 

Since the strategy sets are non-empty, compact, and convex, and as the utility functions \eqref{eq:utility} are continuous and concave in $x_i$, the unregulated IDS game will always have at least one Nash equilibrium (\cite[Proposition 8.D.3]{mas95}).  
{\raggedleft \hfill ${\blacksquare}$}

\subsection*{Proof of Theorem 2}

Consider the optimal security investment profile $\mathbf{x}^*$ in the solution to the centralized problem \eqref{eq:Pcd}. Our goal is to show that there indeed exists a Nash equilibrium $\mathbf{m}^*$ of the mechanism for which $\mathbf{\hat{x}}(\mathbf{m}^*)=\mathbf{x}^*$.  

We start by showing that given the investment profile $\mathbf{x}^*$, it is possible to find a vector of personalized (Lindhal) prices $\mathbf{l}_i^*$, for each $i$, such that,
\begin{eqnarray}
\arg \min_{\mathbf{x}\succeq 0} \quad  g_i(\mathbf{x}) + {\mathbf{l}_i^*}^T \mathbf{x} = \mathbf{x}^*~.
\label{eq:pf2-1}
\end{eqnarray}
First, we know that since $\mathbf{x}^*$ is the solution to problem \eqref{eq:Pcd}, it should satisfy the following KKT conditions, where $\boldsymbol{\lambda}_i \in \mathbb{R}^N_+, \ \forall i$: 
\begin{eqnarray}
\sum_{i=1}^N (\nabla g_i(\mathbf{x}^*) - \boldsymbol{\lambda}_i^T) = \mathbf{0} ~, \notag\\
\boldsymbol{\lambda}_i^T \mathbf{x}^* = 0 \quad \forall i~. 
\label{eq:pf2-2}
\end{eqnarray}
Choose $\mathbf{l}_i^* = - \nabla g_i(\mathbf{x}^*) + \boldsymbol{\lambda}_i^T$. Then, 
\begin{eqnarray}
\mathbf{l}_i^* + \nabla g_i(\mathbf{x}^*) - \boldsymbol{\lambda}_i^T = \mathbf{0}~. 
\label{eq:pf2-3}
\end{eqnarray}
Equations \eqref{eq:pf2-2} and \eqref{eq:pf2-3} together are the KKT conditions for the convex optimization problem: 
\begin{eqnarray}
\min_{\mathbf{x}\succeq 0} \quad g_i(\mathbf{x}) + {\mathbf{l}_i^*}^T \mathbf{x}~. 
\label{eq:pf2-4}
\end{eqnarray}
The KKT conditions are necessary and sufficient for finding the optimal solution to the convex optimization problem \eqref{eq:pf2-4}, and thus we have found the personalized prices satisfying \eqref{eq:pf2-1}. 

We now proceed to finding a Nash equilibrium $\mathbf{m}^*$ implementing the socially optimal solution $\mathbf{x}^*$.  
Consider the message profiles $\mathbf{m}^*_i = (\boldsymbol{\pi}_i^*, \mathbf{x}_i^*)$, for which $\mathbf{x}_i^* = \mathbf{x}^*$, and the price vector proposals $\boldsymbol{\pi}_i^*$ are found from the recursive equations: 
\begin{eqnarray}
\boldsymbol{\pi}_{i+1}^* - \boldsymbol{\pi}_{i+2}^* = \mathbf{l}_i^*, \quad \forall i~.
\label{eq:pf2-5}
\end{eqnarray}
Here, $\mathbf{l}_i^*$ are the personalized prices defined at the beginning of the proof. The set of equations \eqref{eq:pf2-5} always has a non-negative set of solutions $\boldsymbol{\pi}_{i}^* \succeq 0, \ \forall i$. This is because starting with a large enough $\boldsymbol{\pi}_{1}^*$, the remaining $\boldsymbol{\pi}_{i}^*$ can be determined using:\footnote{In \eqref{eq:pf2-6}, $\mathbf{l}_{0}^*$ is interpreted as $\mathbf{l}_{N}^*$.}   
\begin{eqnarray}
\boldsymbol{\pi}_{i}^* = \boldsymbol{\pi}_{i-1}^* - \mathbf{l}_{i-2}^*, \quad \forall i\geq 2~.
\label{eq:pf2-6}
\end{eqnarray}

Now, first note that by \eqref{eq:pf2-4}, for all choices of $\mathbf{x}\succeq 0$, and all users $i$, we have: 
\begin{eqnarray}
g_i(\mathbf{{x}^*}) + {\mathbf{l}_i^*}^T \mathbf{{x}^*} 
\leq g_i(\mathbf{x}) + {\mathbf{l}_i^*}^T \mathbf{x}~.
\label{eq:pf2-7}
\end{eqnarray}
Particularly, if we pick $\mathbf{x} = \frac{\mathbf{x}_i + \sum_{j\neq i} \mathbf{x}_j^*}{N}$, 
\begin{eqnarray}
&& g_i(\mathbf{{x}^*}) + {\mathbf{l}_i^*}^T \mathbf{{x}^*}\notag\\ 
&\leq & g_i(\frac{\mathbf{x}_i + \sum_{j\neq i} \mathbf{x}_j^*}{N}) + {\mathbf{l}_i^*}^T \frac{\mathbf{x}_i + \sum_{j\neq i} \mathbf{x}_j^*}{N}~.
\label{eq:pf2-8}
\end{eqnarray}
Also, since by construction $\mathbf{x}_i^*=\mathbf{x}_{i+1}^*, \ \forall i$, the inequality is preserved for any choice of $\boldsymbol{\pi}_i\succeq 0$, when the two additional tax terms are added in as follows: 
\begin{eqnarray}
&& g_i(\mathbf{{x}^*}) + {\mathbf{l}_i^*}^T \mathbf{{x}^*} + (\mathbf{x}^*_{i} - \mathbf{x}^*_{i+1})^T \text{diag}(\boldsymbol{\pi}^*_i)(\mathbf{x}^*_{i} - \mathbf{x}^*_{i+1})\notag\\
&& ~~~~ - (\mathbf{x}^*_{i+1} - \mathbf{x}^*_{i+2})^T \text{diag}(\boldsymbol{\pi}^*_{i+1})(\mathbf{x}^*_{i+1} - \mathbf{x}^*_{i+2})\notag\\
&\leq& g_i(\frac{\mathbf{x}_i + \sum_{j\neq i} \mathbf{x}_j^*}{N}) + {\mathbf{l}_i^*}^T \frac{\mathbf{x}_i + \sum_{j\neq i} \mathbf{x}_j^*}{N}\notag\\
&& ~~~~ + (\mathbf{x}_{i} - \mathbf{x}^*_{i+1})^T \text{diag}(\boldsymbol{\pi}_i)(\mathbf{x}_{i} - \mathbf{x}^*_{i+1})\notag\\ 
&& ~~~~ - (\mathbf{x}^*_{i+1} - \mathbf{x}^*_{i+2})^T \text{diag}(\boldsymbol{\pi}^*_{i+1})(\mathbf{x}^*_{i+1} - \mathbf{x}^*_{i+2})~.
\label{eq:pf2-9}
\end{eqnarray}
Equation \eqref{eq:pf2-9} can be more concisely written as: 
\begin{eqnarray}
u_i(h({m}^*_i, \mathbf{{m}^*}_{-i})) \geq u_i(h({m}_i, {\mathbf{{m}}^*}_{-i}))~,\notag\\
~~~~~~~~~~~~~~ \forall m_i = (\boldsymbol{\pi}_i, \mathbf{x}_i),\ \forall i~.
\label{eq:pf2-10}
\end{eqnarray}
We conclude that the messages $\mathbf{m}_i^* = (\boldsymbol{\pi}_i^*, \mathbf{x}^*)$ constitute an NE of the proposed mechanism. In other words, the message exchange process will indeed have an NE which implements the socially optimal solution of problem \eqref{eq:Pcd}. 
{\raggedleft \hfill ${\blacksquare}$}